\documentclass[10pt,conference]{IEEEtran}
\IEEEoverridecommandlockouts

\usepackage[T1]{fontenc}

\usepackage{amsmath,amssymb,amsfonts}
\usepackage{xcolor}
\usepackage{graphicx}
\usepackage[numbers]{natbib}
\usepackage[hyphens]{url}
\usepackage{hyperref}
\usepackage{soul}
\usepackage{subcaption}
\usepackage{siunitx}
\usepackage{float}
\usepackage{booktabs}
\usepackage{multirow}
\usepackage{tablefootnote}
\usepackage[linesnumbered,ruled,vlined]{algorithm2e}
\usepackage{longtable}
\usepackage{array}   
\usepackage{comment}

\usepackage{tikz}
\usetikzlibrary{arrows.meta}
\usepackage{xcolor}
\definecolor{navyc}{HTML}{355C7D}   
\definecolor{terrc}{HTML}{C47A5A}   
\definecolor{staedge}{HTML}{2F3B47}
\definecolor{linkc}{HTML}{8A97A4}
\definecolor{ringc}{HTML}{C4CDD6}

\title{Is Collision-Free Backoff Worth It in Wi-Fi?}

\author{Mohammad Yousefi, Francesc Wilhelmi, Boris Bellalta \\[2pt]
  \textit{Wireless Networking - Universitat Pompeu Fabra, Barcelona}\\
  \texttt{\{mohammad.yousefi, francisco.wilhelmi, boris.bellalta\}@upf.edu}}
  
  \date{}

\begin{document}

\maketitle

\begin{abstract}
The Distributed Coordination Function (DCF)---the underlying channel access protocol in Wi-Fi, based on Carrier Sense Multiple Access with Collision Avoidance (CSMA/CA) and Binary Exponential Backoff (BEB)---is simple and effective, but its performance can degrade with increasing contention due to collisions. Collision-free backoff algorithms replace randomization with deterministic channel access, potentially improving efficiency and predictability. But are collision-free methods, with all their implications in terms of protocol design, sufficient to improve Wi-Fi performance under realistic non-full-buffer traffic with both uplink and downlink transmissions? We investigate this question by comparing three collision-free algorithms, CSMA/ECA (ECA), CSMA/E2CA (E2CA), and deterministic backoff (DetBO), with standard and single-stage BEB. Using system-level simulations, we evaluate full-buffer traffic, non-full-buffer ON/OFF traffic with both uplink and downlink transmissions, and coexistence with legacy BEB stations. Under full-buffer traffic, collision-free operation provides only modest throughput gains, up to 6.3\%. With non-full-buffer traffic, delay is governed primarily by the contention window (CW) rather than by collision avoidance, and collision-free access can even degrade performance under downlink-heavy traffic. Stations fall back to a random backoff at least 93\% of the time, so a collision-free schedule rarely forms outside the AP.
\end{abstract}



\section{Introduction}
\label{sec:intro}

Wi-Fi~\cite{geraci2026wi} operates in unlicensed spectrum, which is open to diverse networks and technologies. To enable coexistence among multiple devices, Wi-Fi channel access is driven by Listen Before Talk (LBT). In particular, Wi-Fi devices employ the Distributed Coordination Function (DCF),\footnote{Although current IEEE~802.11 networks use Enhanced Distributed Channel Access (EDCA), we adopt DCF with a single traffic type for simplicity.} which combines Carrier Sense Multiple Access with Collision Avoidance (CSMA/CA) and Binary Exponential Backoff (BEB). Before transmitting over the wireless medium, a station performs Clear Channel Assessment (CCA) checks to ensure that the channel is idle. In parallel, it selects a random backoff from a contention window (CW) and decrements the counter while the channel remains idle. Randomization requires no coordination between stations and provides long-term fairness, which is why it has remained the basis of Wi-Fi channel access. The trade-off is that collisions become inevitable when multiple stations select the same transmission slot. With BEB, stations double their CW after a collision, reducing the likelihood of subsequent collisions but increasing the delay before retransmission and potentially creating a long tail in the packet delay distribution.

At the same time, applications targeted by new Wi-Fi generations, such as IEEE~802.11bn, impose stringent latency and reliability requirements. For example, interactive cloud virtual reality requires network delays between 3 and 10\,ms with reliability greater than 99.9\%, while tactile and haptic traffic requires delays between 1 and 5\,ms with reliability greater than 99.999\%~\cite{adame2021tsn}. Meeting these requirements remains challenging, particularly in networks that rely on randomized channel access~\cite{carrascosa2024performance}.

To prevent collisions inherent to randomized access, several works propose making channel access deterministic. After a successful transmission, instead of drawing a new random value, a station selects a deterministic backoff value. Once stations settle on different transmission schedules, their transmissions can converge to a repeating collision-free cycle in which each station transmits once per cycle. Unlike centralized schedulers already available in the standard, such as Point Coordination Function (PCF) and Restricted Target Wake Time (R-TWT)~\cite{adame2021tsn}, these schemes require neither a coordinator nor explicit signaling, since each station determines its backoff from local observations. While differing in how the deterministic backoff is selected and how collisions are handled, prominent mechanisms following this principle include CSMA with Enhanced Collision Avoidance (CSMA/ECA, hereafter ECA)~\cite{barcelo2008lbeb,barcelo2011towards}; CSMA/ECA using two consecutive deterministic backoffs (CSMA/E2CA, hereafter E2CA)~\cite{barcelo2011towards,sanabria2017high}; and deterministic backoff (DetBO), discussed in IEEE~802.11bn standardization~\cite{wentink2018det,wentink2024det}.

However, whether eliminating collisions justifies the rigid constraints of deterministic access in modern Wi-Fi networks remains debatable. First, imposing a rigid transmission schedule on an uncoordinated set of Wi-Fi devices can be counterproductive: each station receives one transmission opportunity per cycle regardless of its traffic load. This can penalize heavily loaded stations, while stations that become idle leave the schedule; when a new packet arrives, they restart the access process with a random backoff. Thus, under non-full-buffer traffic, the deterministic schedule can be repeatedly disrupted by stations becoming active and idle. Second, collisions may no longer be particularly costly. Request to Send (RTS) / Clear to Send (CTS) confines most of a collision to a short control exchange, while Aggregate Medium Access Control (MAC) Protocol Data Unit (A-MPDU) frame aggregation amortizes channel-access overhead across many frames. Consequently, the elimination of collisions may provide limited benefits when their cost is small compared to the delays introduced by a more conservative or rigid access pattern.

This paper investigates whether collision-free backoff provides tangible benefits in modern Wi-Fi, considering both full-buffer and realistic traffic conditions. To answer this, we compare standard BEB, single-stage BEB, ECA, E2CA, and DetBO. Our main contributions are threefold. First, we describe relevant collision-free backoff schemes proposed for Wi-Fi. Second, we provide an IEEE~802.11 simulator that implements the considered backoff schemes under a common set of assumptions. Third, we comprehensively evaluate their performance with both uplink and downlink transmissions between stations and the Access Point (AP), considering different traffic loads, full-buffer and non-full-buffer operation, and coexistence with legacy BEB stations.


\section{Backoff Schemes}
\label{sec:background}

All schemes follow the IEEE 802.11 DCF operation. A station with a frame to transmit draws an initial backoff counter, decrements it once per idle slot, freezes it during neighboring transmissions, and transmits when the counter reaches zero. A station that empties its buffer leaves contention and clears its backoff state, resetting its contention window to $\mathrm{CW}_{\min}$ and discarding any deterministic backoff its scheme was holding. A station in this cleared state re-enters contention with a random backoff drawn uniformly from $[0,\mathrm{CW}_{\min}]$, regardless of the scheme. The considered backoff schemes therefore differ only in how the backoff is selected after a transmission while the station remains active, and in how collisions are handled.

\subsection{Binary Exponential Backoff}

In BEB, every backoff is random~\cite{ieee80211}. A station draws its value uniformly from $[0,\mathrm{CW}]$, where $\mathrm{CW}=2^{k}(\mathrm{CW}{\min}+1)-1$ and the backoff stage $k$ is the number of successive collisions, capped at six so that $\mathrm{CW}$ never exceeds $\mathrm{CW}{\max}=1023$. A successful transmission resets $k$ to zero, restoring $\mathrm{CW}=\mathrm{CW}_{\min}$. Randomization prevents persistent collisions but also prevents convergence to a collision-free schedule, so the collision probability remains nonzero under saturation~\cite{bianchi2000}.

We consider two variants. Multi-stage BEB is the standard configuration described above. Single-stage BEB disables the exponential stages, keeping $k$ at zero so that every backoff is drawn from $[0,\mathrm{CW}_{\min}]$.


\subsection{ECA}
\label{sec:eca}

ECA replaces the random post-success backoff with a deterministic value
$B_d=\left\lceil\frac{\mathrm{CW}{\min}}{2}\right\rceil-1$, while retaining BEB after collisions~\cite{barcelo2008lbeb,barcelo2011towards}. Once stations occupy different positions in the resulting cycle, transmissions become collision-free. With $\mathrm{CW}{\min}=15$, the deterministic cycle contains $B_d+1=8$ slots and therefore supports at most eight contenders. 

E2CA extends this idea by keeping the deterministic backoff after the first collision following a success, reverting to BEB only after a second consecutive collision~\cite{barcelo2011towards}. This additional tolerance improves robustness against transient collisions while preserving the collision-free schedule. 

\subsection{Deterministic Backoff}

The deterministic backoff proposal~\cite{wentink2018det}, currently under discussion in IEEE 802.11bn~\cite{wentink2024det} and also extended to coexist with New Radio Unlicensed~\cite{tinnirello2026det}, adapts the deterministic backoff to the observed channel activity. Following~\cite{tinnirello2026det}, after a successful transmission, a station selects the deterministic backoff $b=\alpha+i$, where $\alpha$ is a fixed offset, set to $\alpha=10$ here, and $i$ is the number of interruptions observed during the preceding backoff countdown. Consecutive busy periods without an intervening idle slot count as a single interruption. Thus, unlike ECA, the deterministic backoff adapts to the degree of contention.

Unlike BEB, DetBO never doubles the CW after collisions. Instead, collision recovery is governed by two parameters, a modulus $m$ and a threshold $\tau$, set here to $m=7$ and $\tau=3$ following the worked example of~\cite{tinnirello2026det}. After each collision, a retransmission counter $r$ is incremented. If $r\bmod m<\tau$, the station keeps its deterministic backoff. Otherwise, it temporarily draws a random backoff from $[0,m-1]$ before returning to deterministic operation~\cite{tinnirello2026det}. With these values, a station keeps its deterministic backoff through the first two consecutive collisions and falls back to a random one after the third.

A station returning from the cleared backoff state has no previous countdown from which to compute $i$ and therefore starts with a random backoff. It switches to deterministic operation after its first successful transmission.


\section{Evaluation Setup}
\label{sec:model}

This section describes the simulation environment used to compare the considered backoff schemes.\footnote{The simulator's open-source code is provided along with the simulation data, figures, and scripts used to reproduce the reported results of this paper: \url{https://github.com/mmd-yousefi/wifi-backoff-comparison}.} We first establish a common scenario with a single Basic Service Set (BSS) in which all schemes operate under identical physical (PHY) and MAC layer parameters, and then evaluate them under progressively more realistic traffic and coexistence conditions. The evaluation is designed to separate the benefits of collision avoidance from those of the CW policy, particularly when stations are not continuously backlogged. Unless otherwise stated, all results are averaged over multiple independent simulation runs using the same network configuration. 

\subsection{System and Traffic Model}

We evaluate the backoff algorithms in a single BSS consisting of one AP and $N$ stations. The AP and all stations use two spatial streams and an 80~MHz channel, and transmit at 20~dBm. All stations are at the same distance $d=4.5$~m from the AP. With a path loss of $69.4$~dB, the resulting received power per stream and 20~MHz channel is $-58.4$~dBm, and every link uses Modulation and Coding Scheme (MCS)~8.\footnote{The MCS determines the transmission time, and thus the absolute throughput and delay, but not the contention behavior. Using the same MCS for all nodes isolates the impact of the backoff algorithm.} The complete set of simulation parameters is given in Table~\ref{tab:params}.

RTS/CTS is enabled, so collisions are limited to the RTS exchange and the following timeout ($154~\mu$s). Successful transmissions can aggregate up to 64 MAC Protocol Data Units (MPDUs), depending on queue occupancy, resulting in transmission durations between 338 and 1378~$\mu$s. This configuration creates a setting in which collisions are relatively inexpensive: RTS/CTS limits the airtime lost to a collision, while A-MPDU aggregation amortizes channel-access overhead across multiple frames.

Unless otherwise stated, traffic follows an ON/OFF model. ON and OFF periods are exponentially distributed with mean durations of 20~ms and 80~ms, respectively. During ON periods, packets arrive according to a Poisson process, whereas no packets are generated during OFF periods. Each node buffers up to $K$ packets. The AP carries all downlink traffic as a single contending node, selecting the destination station uniformly at random for each Transmission Opportunity (TXOP). 

We set $N=7$, giving at most eight contenders with the AP. This is exactly the capacity of the deterministic cycle in ECA and E2CA, so neither scheme is limited by its cycle length. The full-buffer scenario is the exception, varying $N$ from 2 to 30 to observe the behavior once this capacity is exceeded.

\begin{table}[t]
\centering
\caption{Simulation parameters.}
\label{tab:params}
\small
\begin{tabular}{ll}
\toprule
Parameter & Value\\
\midrule
\multicolumn{2}{l}{\textbf{PHY}}\\
Channel bandwidth         & 80\,MHz\\
Spatial streams           & 2\\
MCS                       & 8 (256-QAM, coding rate $3/4$)\\
Transmit power            & 20\,dBm\\
Distance, $d$      & 4.5\,m\\
Slot time                 & 9\,$\mu$s\\
SIFS / DIFS               & 16 / 34~$\mu$s\\
\midrule
\multicolumn{2}{l}{\textbf{MAC}}\\
CW$_{\min}$ / CW$_{\max}$ & 15 / 1023\\
RTS/CTS                   & Enabled\\
Max.\ A-MPDU size         & 64 MPDUs\\
MPDU payload              & 1500\,B\\
Buffer size, $K$           & 1000 MPDUs\\
\midrule
\multicolumn{2}{l}{\textbf{Traffic}}\\
Model                     & ON/OFF\\
Mean ON / OFF             & 20 / 80\,ms\\
\midrule
\multicolumn{2}{l}{\textbf{Simulation}}\\
Number of stations, $N$              & 7 (2--30, full buffer)\\
Simulation time                  & 20\,s (30\,s, full buffer)\\
Random seeds              & 10\\
\bottomrule
\end{tabular}
\end{table}

\subsection{Scenarios}
\label{sec:model:scenarios}

We consider three scenarios:
\begin{itemize}
\item \emph{Full-buffer.} Every node is continuously backlogged. We vary the number of contenders and measure the aggregate throughput.
\item \emph{Non-full-buffer.} Nodes follow the ON/OFF traffic model. We consider three traffic configurations: uplink-only, with 100\% of the traffic generated by stations; balanced, with 50\% uplink and 50\% downlink traffic; and downlink-heavy, with 20\% uplink and 80\% downlink traffic. For each configuration, we sweep the offered load.
\item \emph{Coexistence.} Collision-free stations coexist with legacy BEB stations at the medium-high load of the balanced traffic configuration while varying their relative proportions.
\end{itemize}

Each scheme reaches saturation at a different offered load, so there is no single saturation point shared by all schemes. To ensure that all schemes are evaluated under comparable non-saturated conditions, we define a reference load $S$ for each traffic configuration as the highest offered load at which every scheme keeps its packet drop rate below 7\% in both directions. Thus, $S$ is determined by the first scheme to reach this threshold. Because the AP carries all downlink traffic through a single queue, it is typically the first bottleneck in scenarios with downlink traffic. As a result, $S$ decreases from 490~Mbps for uplink-only traffic to 160~Mbps when 80\% of the offered traffic is downlink. The first scheme to reach the threshold is single-stage BEB for the uplink-only and balanced cases, and DetBO for the downlink-heavy case (Table~\ref{tab:loads}).

\begin{table}[t]
\centering
\caption{Offered load (Mbps) at each load level, as fractions of the saturation
load $S$, which is set by the first scheme to saturate.
}
\label{tab:loads}
\footnotesize
\setlength{\tabcolsep}{3pt}
\begin{tabular}{l|c|cccc}
\toprule
Scenario & $S$ & Low & Low-Med & Med-High & High\\
 &  & ($0.2S$) & ($0.4S$) & ($0.7S$) & ($0.9S$)\\
\midrule
Uplink-only (100/0)  & 490 &  98 & 196 & 343 & 441\\
Balanced (50/50) & 225 &  45 &  90 & 158 & 203\\
Downlink-heavy (20/80) & 160 &  32 &  64 & 112 & 144\\
\bottomrule
\end{tabular}
\end{table}

\subsection{Metrics}

We report four metrics, averaged over 10 runs with 95\% confidence intervals. All metrics are computed over the entire simulation, including its initial seconds, since each node starts in the stationary state of its ON/OFF process. The conditional collision probability is the fraction of transmission attempts that collide,
\begin{equation}
p = \frac{\sum_i C_i}{\sum_i A_i},
\end{equation}
where $A_i$ and $C_i$ are the transmission attempts and collisions of node $i$, respectively. With RTS/CTS, an attempt corresponds to an RTS transmission.

The aggregate throughput is the delivered payload per unit time,
\begin{equation}
S_{\mathrm{th}} = \frac{L\sum_j k_j}{T},
\end{equation}
where $k_j$ is the number of MPDUs in the $j$-th successful TXOP, $L$ is the MPDU payload, and $T$ is the run duration.

The packet delay is measured from packet generation to successful delivery,
\begin{equation}
D_n = t^{\mathrm{del}}_n - t^{\mathrm{gen}}_n,
\end{equation}
and we report its mean and 99th percentile over successfully delivered MPDUs. Dropped packets are excluded from the delay statistics. A packet is dropped only when it arrives at a full buffer, since there is no retry limit and a frame is retransmitted until it is delivered.


\section{Results}
\label{sec:results}

This section compares the five backoff schemes in terms of throughput, conditional collision probability, and packet delay. We evaluate them under the traffic and coexistence scenarios described in Section~\ref{sec:model:scenarios}. 

\subsection{Full-buffer Traffic}
\label{sec:results:saturated}

We first consider full-buffer traffic, where every node always has uplink traffic to transmit to the AP. This represents the operating point that most favors collision-free schemes, since contention is highest and deterministic schedules, once formed, are never disrupted. 

\begin{figure}[t]
  \centering
  \includegraphics[width=\columnwidth]{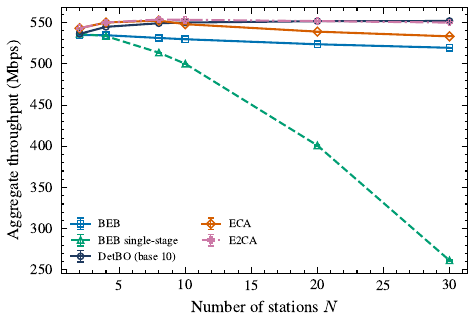}
  \caption{Aggregate throughput versus the number of contenders under full-buffer traffic.}
  \label{fig:saturated}
\end{figure}

The results are shown in Fig.~\ref{fig:saturated}, which reports the aggregate throughput of the five schemes against the number of contenders $N\in \{2, ..., 30\}$. BEB, ECA, E2CA, and DetBO obtain between 519 and 554\,Mbps over the entire range of contenders, whereas single-stage BEB decreases from 536 to 262\,Mbps. The limited throughput gain of collision-free schemes compared to BEB stems from the low cost of collisions. With RTS/CTS enabled, a collision occupies only the short RTS exchange, about one ninth of the airtime of a successful transmission, while each successful transmission carries up to 64 aggregated MPDUs. Even BEB therefore sustains 519\,Mbps despite a conditional collision probability of 0.53 at $N=30$. Single-stage BEB is the exception, because its CW never grows. With 30 contenders, each of them collides on 97\% of its transmission attempts, and although each collision is cheap, there are enough of them to fill about half of the channel time. Its throughput falls to 262\,Mbps as a result.

The collision probabilities of the collision-free schemes differ substantially. DetBO is the only scheme that converges to fully collision-free operation for every $N$ in the range, because its deterministic backoff adapts to the observed contention. In contrast, ECA and E2CA are limited by their fixed eight-slot schedule. Beyond eight contenders, distinct transmission opportunities can no longer be accommodated, causing their collision probabilities to increase to 0.37 and 0.10, respectively. E2CA degrades more gracefully because the deterministic schedule is preserved after an isolated collision, allowing faster re-convergence.

\vspace{0.1cm}
\noindent \textbf{Takeaway:} Under full-buffer traffic, collision-free operation provides only modest throughput gains compared to BEB because RTS/CTS and A-MPDU aggregation make collisions inexpensive. DetBO and E2CA both outperform BEB by about 6\% at $N=30$, as their deterministic backoff keeps a near-collision-free schedule. Single-stage BEB performs worst, confirming that a bounded CW alone is insufficient when the number of contenders is high.

\subsection{Non-full-buffer Traffic}
\label{sec:results:nonsaturated}

We now consider non-full-buffer traffic with one AP and seven stations, where every node follows the ON/OFF traffic model of Section~\ref{sec:model}. We evaluate three traffic configurations combining uplink and downlink transmissions. The number of contenders is $N=7$ for uplink-only traffic and $N=8$ when both uplink and downlink transmissions are present.

\subsubsection{100\% Uplink}
\label{sec:results:ul}

We first consider uplink-only traffic, using this scenario to directly compare the different backoff schemes. 

\begin{figure*}[t]
  \centering
  \subfloat[Delay]{%
    \includegraphics[width=0.48\textwidth]{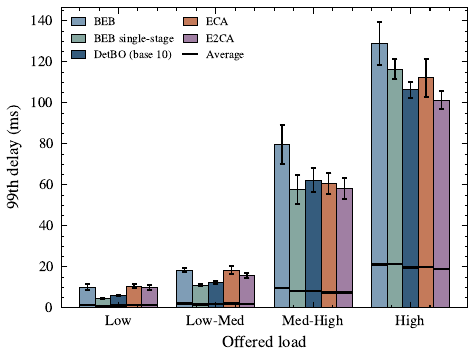}%
    \label{fig:ul:latency}}
  \hfil
  \subfloat[Collision probability]{%
    \includegraphics[width=0.48\textwidth]{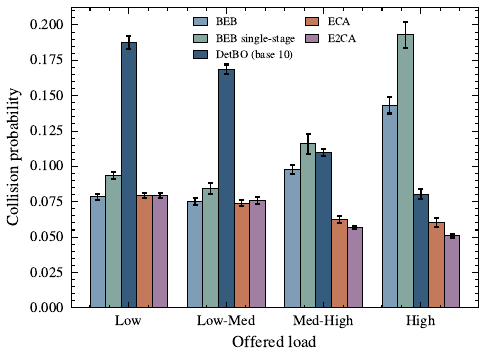}%
    \label{fig:ul:collisions}}
  \caption{100\% Uplink scenario under non-full-buffer traffic: (a) mean and 99th-percentile packet delay and (b) conditional collision probability versus the aggregate offered load.}
  \label{fig:ul}
\end{figure*}

Fig.~\ref{fig:ul:latency} shows the mean and 99th-percentile uplink delay at the four load levels considered. As shown, the average delay remains nearly identical across all schemes, reaching 18.8--21.2\,ms at the highest load. The differences between schemes are much smaller than in the tail, indicating that the average delay is dominated by queueing rather than by the backoff algorithm. The main differences appear in the tail delay. At the highest load, the 99th percentile ranges from 101 to 129\,ms. E2CA and DetBO achieve the shortest tails, 101\,ms and 106\,ms, followed by ECA at 112\,ms. Single-stage BEB reaches 116\,ms even though its collision probability, 0.193, is the highest of all schemes, because its bounded window keeps each retry short. Standard BEB has the largest tail, 129\,ms, because it combines frequent collisions with immediate window growth.

Fig.~\ref{fig:ul:collisions} shows the conditional collision probability of each scheme against the offered load. These values explain the trends above. At the two lowest loads, DetBO exhibits the highest collision probability (0.19 and 0.17, compared to about 0.08 for the other schemes). After a collision, DetBO initially reuses its deterministic backoff $b=\alpha+i$ instead of selecting a new random value, where $i$ counts the busy-period interruptions observed during the preceding countdown. At low loads, the channel is mostly idle, so $i$ is almost always zero and $b$ collapses to the fixed offset $\alpha$ for every station. Two stations that have just collided therefore select the same value again and collide again. As the offered load rises, each station observes a different number of interruptions, $i$ spreads out across stations, and colliders separate after the first attempt. Consequently, DetBO's collision probability falls to 0.081 at the highest load, while BEB and single-stage BEB increase to 0.143 and 0.193, respectively.

\vspace{0.1cm}
\noindent \textbf{Takeaway:} Under non-full-buffer uplink traffic, collision avoidance has little impact on the average delay. The tail delay is more sensitive to the backoff scheme, and a fixed CW that never doubles compensates for a high collision rate, as single-stage BEB shows with the highest collision probability but not the longest tail.

\subsubsection{50\% Uplink -- 50\% Downlink}
\label{sec:results:5050}

We now consider balanced traffic, where the AP carries half of the total offered load while contending on equal terms with the stations. This scenario highlights how each backoff algorithm treats a node that is substantially more loaded than its competitors. We therefore report uplink and downlink delays separately. 

\begin{figure*}[t]
  \centering
  \subfloat[Uplink delay]{%
    \includegraphics[width=0.48\textwidth]{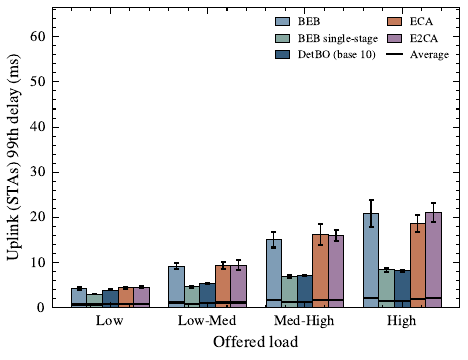}%
    \label{fig:5050:ul}}
  \hfil
  \subfloat[Downlink delay]{%
    \includegraphics[width=0.48\textwidth]{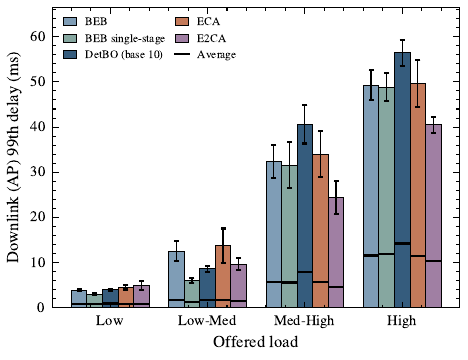}%
    \label{fig:5050:dl}}
  \caption{Balanced traffic (50\% uplink, 50\% downlink) under non-full-buffer traffic: mean and 99th-percentile packet delay of (a) uplink stations and (b) AP downlink transmissions versus the aggregate offered load.}
  \label{fig:5050}
\end{figure*}

Fig.~\ref{fig:5050} shows the delay of the stations and of the AP separately, at the four load levels. As shown, uplink and downlink traffic experience similar delays at low load but diverge as the offered load increases. At the highest load, the 99th-percentile downlink delay reaches 41--56\,ms, whereas the uplink remains below 22\,ms. As observed in Section~\ref{sec:results:ul}, DetBO and single-stage BEB, whose CWs do not grow after a collision, achieve the shortest tails (8.2 and 8.4\,ms), whereas BEB, ECA, and E2CA reach 19--21\,ms due to exponential backoff. The downlink reveals a different behavior. The AP becomes the bottleneck because it serves half of the traffic through a single transmission queue while obtaining channel access no more frequently than any individual station. DetBO, which achieves the lowest uplink delay, now exhibits the largest downlink tail (56.4\,ms). The reason is that the AP and the stations do not experience the same contention while they count down, so although both follow the same deterministic rule, their backoffs settle at different values. The AP's queue drains between bursts, but since it carries half of the traffic, it usually still has packets waiting from one transmission to the next. A station, in contrast, empties its buffer and rejoins with a shorter random backoff. The AP therefore counts down for longer and senses more busy periods, and its deterministic backoff grows accordingly, averaging 11.4 slots against 10.1 for the stations. As a result, the most loaded node is the one that waits the longest, and its queue drains slowly. Under the random schemes, the AP draws its backoff from the same window as the stations, so it carries no systematic penalty and clears its backlog more quickly.

The collision probability, which we report in the text only for the sake of space, does not explain the downlink delay. DetBO collides about twice as often as BEB (0.201 versus 0.098) and performs worst, whereas ECA collides almost identically to BEB (0.098 for both) and achieves essentially the same delay. E2CA provides the shortest downlink tail (40.5\,ms) despite exhibiting the same collision probability as BEB (0.098). Its advantage comes from delaying CW growth after an isolated collision, preventing the AP from repeatedly entering large backoff stages.

\vspace{0.1cm}
\noindent \textbf{Takeaway:} Under 50\% Uplink and 50\% Downlink traffic, the AP becomes the bottleneck. Deterministic scheduling gives it a longer backoff than the stations it serves, so the busiest node obtains the fewest transmission opportunities. The downlink delay is therefore dominated by how each backoff algorithm manages the CW of this node, not by its collision probability. DetBO gives the AP the longest delay because it grows the AP's backoff with the channel activity it senses, whereas E2CA gives the shortest because it does not grow the CW after an isolated collision.

\subsubsection{20\% Uplink -- 80\% Downlink}
\label{sec:results:2080}

\begin{figure*}[t]
  \centering
  \subfloat[Uplink delay]{%
    \includegraphics[width=0.48\textwidth]{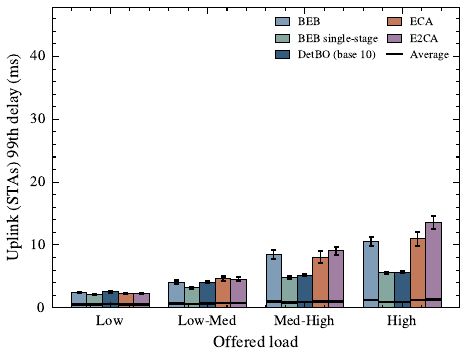}%
    \label{fig:2080:ul}}
  \hfil
  \subfloat[Downlink delay]{%
    \includegraphics[width=0.48\textwidth]{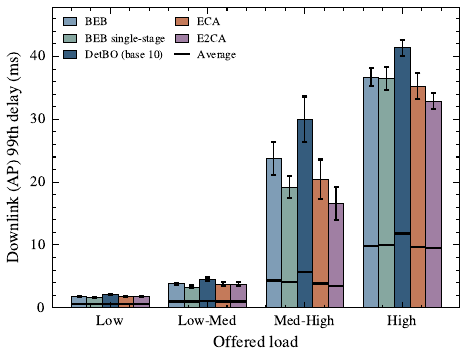}%
    \label{fig:2080:dl}}
  \caption{Downlink-heavy traffic (20\% uplink, 80\% downlink) under non-full-buffer traffic: mean and 99th-percentile packet delay of (a) uplink stations and (b) AP downlink transmissions versus the aggregate offered load.}
  \label{fig:2080}
\end{figure*}

The final scenario is downlink-heavy traffic, in which the AP carries 80\% of the total network load. This closely resembles typical Wi-Fi deployments. The asymmetry of the 50-50\% case is sharper here, since the AP now carries 115\,Mbps against 4\,Mbps per station. We use this sharper asymmetry to test whether the findings of Section~\ref{sec:results:5050} still hold, and to measure directly how much of the deterministic schedule survives at each node. 

Fig.~\ref{fig:2080} shows the delay of the stations and of the AP separately, at the four load levels. At the highest load the 99th-percentile downlink delay reaches 32.9--41.4\,ms, about three times the uplink, which stays below 14\,ms for every scheme. Every conclusion of the 50-50\% case holds unchanged. On the uplink, DetBO and single-stage BEB, whose CWs do not grow after a collision, achieve the shortest tails (5.6--5.7\,ms), while BEB, ECA, and E2CA reach 10--14\,ms due to exponential backoff. On the downlink, DetBO again produces the longest tail (30 and 41.4\,ms at the two highest loads) and E2CA the shortest (16.6 and 32.9\,ms). At the highest load DetBO collides about twice as often as BEB (0.164 versus 0.077), yet its stations still obtain one of the shortest uplink delays. DetBO also drops the most packets, 2.5\% against 1.2\% for E2CA, all at the AP, whose slower queue occasionally fills. These packets never enter the delay statistics, so DetBO's downlink delay is even worse than the figure shows.

To see how much of the deterministic schedule actually survives, we count, at the highest load, how many of the deterministic backoffs each node takes fall back to a random one before being used. This happens for two reasons. Either the node collides and the recovery rule replaces the way the next backoff is chosen, or its buffer empties and it later rejoins contention with a random backoff. Table~\ref{tab:fallback} reports the share of fallbacks and their cause, for the stations and for the AP. BEB and single-stage BEB draw every backoff at random, so the metric does not exist for them.

\begin{table}[t]
\centering
\caption{Share of deterministic backoffs that fall back to a random one, and the cause of the fallback, under 20\% Uplink -- 80\% Downlink traffic at the highest load.}
\label{tab:fallback}
\small
\begin{tabular}{llrrr}
\toprule
Scheme & Node & Fallback & Going idle & Collision \\
\midrule
ECA   & Stations & 100.0\% & 100.0\% & 0.0\% \\
      & AP  & 28.7\%  & 20.6\%  & 8.1\% \\
E2CA  & Stations & 100.0\% & 100.0\% & 0.0\% \\
      & AP  & 19.2\%  & 18.3\%  & 0.9\% \\
DetBO & Stations & 93.5\%  & 88.5\%  & 5.0\% \\
      & AP  & 19.7\%  & 17.3\%  & 2.4\% \\
\bottomrule
\end{tabular}
\end{table}

The stations practically never maintain their deterministic schedules. ECA and E2CA lose essentially all of their deterministic backoffs (100\%), while DetBO retains only 6.5\%. In all cases, schedules are lost mainly because stations become idle rather than because of collisions. DetBO achieves slightly higher retention because its rule preserves the deterministic value after the first two consecutive collisions; however, most of these retained values correspond to re-attempts of a slot that has just failed. In contrast, the AP is the only node that consistently maintains a schedule, retaining between 71\% and 81\% of its deterministic backoffs, because it goes idle less often than a station. Thus, collision-free operation is effectively sustained only by the AP, while under DetBO, this is precisely the node that is penalized by the deterministic schedule.

\vspace{0.1cm}
\noindent \textbf{Takeaway:} Increasing the downlink share changes none of the conclusions, but it reveals where the deterministic schedule actually survives. Stations abandon their deterministic backoff before using it at least 93\% of the time, almost always because their buffers empty, so collision-free operation barely exists on the uplink. The AP empties its buffer less often and is therefore the only node that sustains a schedule. Under DetBO, however, sustaining that schedule leaves the AP with a longer backoff than the stations, and the node carrying most of the traffic waits the longest.

\subsection{Coexistence with Legacy Stations}
\label{sec:results:coexistence}

We finally consider a heterogeneous deployment in which collision-free stations coexist with legacy BEB stations. Starting from the balanced scenario at the medium-high load (158\,Mbps) with all eight contenders running ECA, E2CA, or DetBO, we progressively replace them with legacy BEB nodes. Each mix is written as the number of collision-free contenders against the number of legacy ones. The AP is one of the eight and keeps its collision-free scheme in every mix but the all-BEB baseline, so 6/2 means five collision-free backoff stations and the AP against two BEB stations.

\begin{figure*}[t]
  \centering
  \subfloat[ECA vs BEB]{%
    \includegraphics[width=0.32\textwidth]{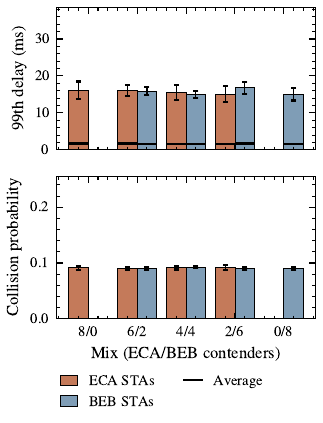}%
    \label{fig:coex:eca}}
  \hfil
  \subfloat[E2CA vs BEB]{%
    \includegraphics[width=0.32\textwidth]{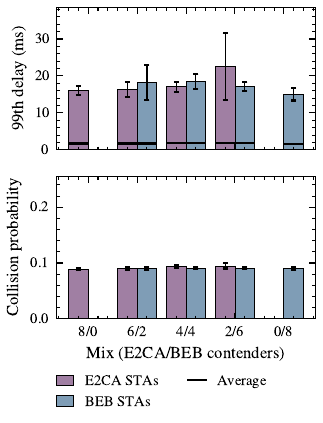}%
    \label{fig:coex:e2ca}}
  \hfil
  \subfloat[DetBO vs BEB]{%
    \includegraphics[width=0.32\textwidth]{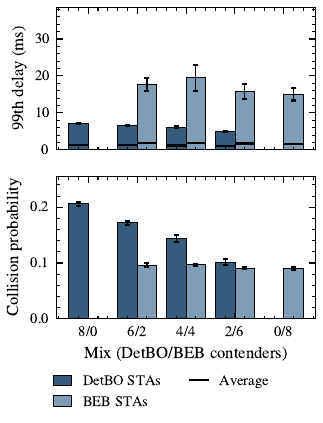}%
    \label{fig:coex:det}}
  \caption{Coexistence with legacy BEB under the balanced medium-high load. Each subfigure compares one collision-free scheme against legacy BEB stations, reporting mean and 99th-percentile delay (top) and conditional collision probability (bottom) versus the mix of collision-free and legacy contenders. All subfigures report the stations only, the AP results are given in the text.}
  \label{fig:coex}
\end{figure*}

Fig.~\ref{fig:coex} compares each scheme against legacy BEB, keeping the collision-free and the legacy stations apart and sharing the all-BEB baseline. The AP is left out of all subfigures, since it carries half the traffic and does not compare with a station, so its results are given in the text. ECA (Fig.~\ref{fig:coex:eca}) and E2CA stations (Fig.~\ref{fig:coex:e2ca}) are statistically indistinguishable from the legacy stations they share the channel with, in both collision probability and delay. DetBO stations (Fig.\ref{fig:coex:det}) collide less often as their proportion decreases, from 0.206 to 0.101, approaching the all-BEB value of 0.090 because repeated deterministic collisions become unlikely when competing with random backoff. Despite this, their 99th-percentile delay remains between 5.0 and 7.1\,ms, compared with 15.9--19.5\,ms for BEB stations. Legacy stations are unaffected: both their collision probability (0.090--0.097) and delay remain statistically unchanged across all coexistence ratios.

The AP exhibits the same behavior as in the previous scenarios. ECA leaves the collision probability and the 99th-percentile delay unchanged (0.112 and 34.0\,ms), while E2CA keeps the AP delay below the BEB baseline (24.4\,ms) despite a similar collision probability (0.108). Running DetBO increases its collision probability to 0.178 and its 99th-percentile delay to 40.6\,ms, compared with 0.112 and 32.3\,ms under the all-BEB baseline. 

Stations repeatedly empty their buffers and rejoin with a random backoff, so the deterministic schedule rarely persists. ECA then behaves like BEB, since any collision returns it to BEB in any case. E2CA differs only in tolerating one collision before doubling, which helps the AP, whose delay is set by how large its CW becomes. DetBO instead grows the AP's backoff with the activity it senses, so the busiest node waits the longest.

\vspace{0.1cm}
\noindent \textbf{Takeaway:} Collision-free schemes can coexist with legacy BEB without degrading legacy performance. DetBO obtains the shortest station delay despite the highest collision probability, because its CW never grows after a collision. ECA and E2CA behave almost identically to BEB because they follow the same CW policy and persistent deterministic schedules rarely form.

\section{Conclusions}
\label{sec:conclusions}

We compared standard BEB, single-stage BEB, ECA, E2CA, and DetBO under full-buffer and non-full-buffer traffic, as well as in coexistence with legacy stations, to assess the benefits of collision-free backoff. The gains are modest and, under downlink-heavy traffic, deterministic scheduling can even degrade performance.

RTS/CTS and A-MPDU aggregation make collisions inexpensive, limiting the throughput gain of collision-free operation under full-buffer traffic to 6.3\%. Under non-full-buffer traffic, the delay tail is governed primarily by the CW policy rather than by collision avoidance, as single-stage BEB shows with the highest collision probability but not the longest tail. Consequently, collision probability alone is a poor performance indicator. The same holds in coexistence with legacy BEB, where collision-free stations cause no measurable degradation, and their own benefit again follows the CW policy rather than collision avoidance.

The reason is that a deterministic schedule rarely forms at all. Stations fall back to a random backoff at least 93\% of the time, almost always because their buffers empty before the deterministic value is ever used, so only the AP sustains a schedule, and that is the node that deterministic access penalizes most. 


\section{Acknowledgments}

This work was supported by the following projects: TRUE Wi-Fi PID2024-155470NB-I00 (MICIU/AEI/10,13039/501100011033/FEDER,UE), ICREA Academia 2024 (00077 AGAUR), and MdM CEX2021-001195-M (MICIU/AEI/10.13039/501100011033). 

\bibliographystyle{IEEEtran}
\bibliography{References}

\end{document}